

\def\NP{Nucl.~Phys.~}
\def\PR{Phys.~Rev.~}

\def\PL{Phys.~Lett.~}

\input phyzzx
\pubnum{37}
\date{February 1992}
\titlepage
\title{Path-Integral for Quantum Tunneling}
\author{Hideaki Aoyama\foot{E-mail address: aoyama@jpnyitp.bitnet}
{\rm and} \ Arihiro M.~Tamra}
\vskip 0.3cm
\centerline{Physics Department, College of Liberal Arts and Sciences}
\centerline{Kyoto University, Kyoto 606, JAPAN}
\vskip 1.5cm
\abstract

Path-integral for theories with degenerate vacua is investigated.
The origin of the non Borel-summability of the perturbation theory
is studied.
A new prescription to deal with small coupling is proposed.
It leads to a series, which at low orders and small coupling
differs from the ordinary perturbative series
by nonperturbative amount, but is Borel-summable.

\vskip 3cm
\endpage

\normalspace

\chapter{Introduction}

The perturbation theory plays major roles in quantum mechanics
and in quantum field theories.
In the latter, it often is the only reliable calculational tool
one has. (Cross sections in the QCD are good examples of this.)
In spite of its importance, the perturbation series one obtains, in most
 cases, has zero convergence radius, due to
the presence of the Dyson\Ref\dyson{F.~J.~Dyson
\journal \PR &85 (52) 631.}
singularities in the complex coupling-constant plane.

Several years ago, many important works appeared on the
convergence property of the perturbation series.\Ref
\zinnbook{Most of the references relevant in this context can be found
in; {\sl \lq\lq Large-Order Behavior of Perturbation Theory"}
(edited by J.~C.~Le Guillou and J.~Zinn-Justin, North-Holland 1990).}
As a result, the perturbation series for many models were found to be
asymptotic series.

If the series is asymptotic, it can be summed
by using the Borel summation method.
An alternative the Pad\'e approximation, which is known
to give convergent results for asymptotic series.\Ref
\simon{S.~Graffi, V.~Grecchi and B.~Simon \journal \PL &32B (70) 631.}
The existence of these summation methods guarantees
the validity of using the first few orders of
the perturbation series, as often done in quantum field theories, such
 as QED.

\REF\brezin{F.~Br\'ezin,
   G.~Parisi and J.~Zinn-Justin \journal\PR &D16 (77) 408.}
\REF\lipatov{L.~N.~Lipatov \journal Sov.~Phys.~JETP &45 (77) 216.}
\REF\zinn{E.~Brezin, J.-C.~Le Guillou and J.~Zinn-Justin \journal\PR
 &15 (77) 1544, 1558.}
\REF\morezinn{J.~Zinn-Justin \journal{J.~Math.~Phys} &22 (81) 511
\journal\NP &B192 (81) 125
\journal\NP &B218 (83) 333
\journal{J.~Math.~Phys} &25 (84) 549.}
\REF\aoyamalast{H.~Aoyama, SLAC preprint, SLAC-PUB-5624/KUCP-35 (August
 1991).}

The situation is completely different when the theory allows tunneling.
In such a theory, the perturbative series
is simply divergent (not Borel-summable).
It has a cut in the physical region of the coupling constant.\refmark
{\brezin - \zinn}
Thus the nonperturbative contribution of the tunneling,
which in Euclidean path-integral formalism manifests as instantons,
needs to be accounted for.
Although much progress was made in this approach,\refmark\morezinn
it is not clear how to deal with the
divergence of the perturbation series to obtain quantitatively
reliable results.

In this paper, we will analyse this problem under a new light.
Namely,
in the path-integral formalism with finite periodicity in
Euclidean time,
we will argue that the path-integral has a singularity
at the zero coupling due to the multiplicity of the classical minima,
which does not exist in theories without tunneling.
Based on this, we propose to re-define the path-integral
so that it is equal to the original path-integral
for small coupling but is free of the singularity.
This path-integral is suitable for the evaluation of the
original path-integral for small but non-zero coupling.
This procedure is presented in the chapter 2.
The next chapter gives the demonstration of the power of
this method for a simple one-dimensional integral.
The chapter 4 gives the analysis of a
quantum mechanical model.
(Part of this analysis was
presented in Ref.\aoyamalast\ by one of the authors.)
The last chapter gives the discussion and the summary.
Some numerical results, which shows the validity of our analysis,
are given in the appendix.

\chapter{General formalism}
\section{The origin of non-summability}

\def\nn{{\cal N}}
\def\zg{{\cal Z} (g)}
\def\intphi{\int {\cal D} \phi}
\def\fn{F_n[\phi]}

Let us take
the partition function in the Euclidean formalism,
$$\zg \equiv \nn \intphi\ e^{-S[\phi,g]}, \eqn\zgdef$$
where $g$ is the coupling constant of the theory, and
$\phi$ represents all the dynamic variable of the theory.
[Without losing generality, we assume hereafter that $\phi$ is real.]
The factor $\nn$ is a normalization factor, which
we will elaborate on later.
We also impose a periodic boundary condition in the
imaginary time $\tau \in [0, \beta]$.
This way, the spectrum of the theory can be obtained from the position
 of the
poles in the Laplace-transform of $\zg$ with respect to $\beta$.

In the ordinary perturbation theory
(which generates Feynman graphs in the quantum field theory),
we first expand the integrand in powers of $g$;
$$e^{-S[\phi,g]} = \sum_{n=0}^{\infty} g^n \fn. \eqn\fndef$$
The perturbative coefficients $z_n$, which we define by
a path-integral of $\fn$,
$$z_n = \nn \intphi\ \fn, \eqn\naivezn$$
gives the perturbative series for the partition function;
$${\cal Z}_{\rm pert}(g) \equiv \sum_{n=0}^\infty  z_n g^n.
 \eqn\naive$$
At the zero coupling, $g=0$, we have a free theory,
and $z_0$ is expressed in terms of the determinant of the free theory
in the usual manner.
However, the small coupling behaviour of ${\cal Z} (g)$
is drastically different from the above:
Let us denote the number of the  degenerate minima by $m$.
Then the functional space of $\phi$ has $m$ points
where the integrand $e^{-S[\phi,g]}$ is maximized.
For small enough $g$, $e^{-S[\phi,g]}$ has $m$ well-separated ``peaks"
at these points.
For $g \rightarrow 0$, peaks are completely isolated from other peaks
and thus each peak contributes a free-theory value.
Therefore we find that
$$\lim_{g \rightarrow 0} {\cal Z} (g) = m z_0 = m {\cal
 Z}(0).\eqn\mtimes$$
This is the discontinuity at the zero coupling.
[The situation is illustrated in Fig.1 for $m=2$.]
This discontinuity is the reason for the fact that
the naive perturbation series \naive\ is divergent (not even
 Borel-summable).

When one considers correlation functions and other observables,
the existence of this discontinuity
is obscured by implicit adjustment of the normalization factor.
However, the discontinuity is there and the perturbative calculation
leads to a divergent result.

\FIG\gap{The singularity at the zero coupling.
The actual calculation is done for
the function $\zg$ for the simple model \simple.}

In order to illustrate these arguments,
let us take a very simple analogue of the theory with tunneling.
[Analysis of the actual quantum theory will be given later.]
It is defined by the following ``action" without any time dependence;
$$S(\phi, g) \equiv {1 \over 2} \phi^2 (1 - g \phi)^2 \eqn\simple$$
The partition function is then a simple one-dimensional integral.
Obviously, for zero coupling,
$${\cal Z} (0) = \nn \int\nolimits_{-\infty}^\infty e^{-{1\over
 2}\phi^2}
  = \nn \sqrt{2 \pi}. \eqn\zerocp$$
On the other hand,
$$\lim_{g \rightarrow 0} {\cal Z} (g) = 2\nn \sqrt{2 \pi},\eqn\twice$$
because of the contribution from the two almost-Gaussian
peaks at $\phi=0$ and $1/g$. (Of course, \undertext{at} $g=0$, the
outer peak is at $\phi = \infty$ and does not contribute to the
 integral.
Thus \zerocp.)
This is the discontinuity mentioned above.
In fact, Fig.\gap\ is the actual numerical plot for this simple model,
where we have chosen that $\nn = 1/(2 \sqrt{2\pi})$.
A simple calculation shows that
$${\cal Z}_{\rm pert}(g) = {1\over 2} + 3 g^2 + 105 g^4 + 6930 g^6 + ...
 ,
     \eqn\ordres$$
while for small $g$, $\zg$ should be expressed as
$$\zg = 1 + \epsilon(g) \quad
(\lim_{g\rightarrow 0}\epsilon(g) = 0).\eqn\shouldbe$$
If one is not careful about the normalization factor $\nn$,
the difference of the first terms 1 and 1/2 goes unnoticed,
and one would ``find" the perturbative result to be
twice as much as the \ordres;
$$\tilde{\cal Z}_{\rm pert}(g) =
1 + 6 g^2 + 210 g^4 + 13860 g^6 + ...,
\eqn\ordresa$$
which satisfies the property \shouldbe.
This is what we mentioned as implicit adjustment of the normalization
 factor
in the above.
Numerically, we find that
the first few orders of the perturbative series \ordresa\
(which, of course, is wrongly normalized)
give excellent results for small couplings,
just as perturbative series in QCD does.
We shall see in the next chapter why this is so,
in spite of the facts that
the whole series is non-summable
and that it has wrong normalization factor.

\section{A prescription for small coupling}
The natural question is then what should be done for the small coupling
approximation.
The general prescription we propose is the following.
Let us assume the degeneracy (and tunneling)
is due to a discrete symmetry of the theory.\foot{
Again, QCD is a good example of this.
For cases where this is not satisfied, there are natural extensions of
the method we present in this paper.}
In such a case, the whole functional-space of $\phi$
can be divided to subspaces, which are mapped into
each other by the symmetry.
A subspace that is not mapped onto itself and covers the whole space
by symmetry transformations is called fundamental region.
[This concept is commonly used in closed-string theories.]
If the functional space is covered by $m$ fundamental regions,
the partition function \zgdef\ for $g \ne 0$ can be written as
\def\ss{{\cal R}}
$$\zg = m \nn \int\nolimits_{\ss} {\cal D}\phi \, e^{-S[\phi, g]},
    \eqn\zdefimp$$
where $\ss$ is the fundamental region that contains the classical
vacuum at $g=0$.
We stress here that the above equality does not hold for $g = 0$.
However, since we are interested in the small but non-zero coupling
case, we need to establish a small coupling approximation of
the right hand side of \zdefimp.
That expression would be regular at $g = 0$, and
does not have the right value of ${\cal Z}(0)$,
since in the limit $g\rightarrow 0$,
the degenerate vacua (other than the perturbative one)
moves out to infinity and the fundamental region $\ss$ expands
to cover the whole space.
In the simple example \simple, the resulting expression would
satisfy the property \shouldbe.

\def\limgzero{\lim_{g \rightarrow 0}}
In other words,
the quantity defined by the right-hand side of \zdefimp\
is regular at $g=0$.
Thus we choose to redefine the partition
function by this quantity.
This way, the observables calculated from the partition function
become free from the zero-coupling singularity
and it should make sense to construct
some kind of the perturbation theory.

We construct this ``improved" perturbation theory
by utilizing the expansion of the integrand \fndef.
Namely, the small coupling approximation we obtain is
$$\zg = \sum_{n=0}^\infty k_n(g) \, g^n, \quad
k_n (g) = m \nn \int\nolimits_{\ss} {\cal D}\phi \, \fn.\eqn\zdefimpp$$
{}From the previous discussion that ${\cal R}$
covers the whole space at zero coupling,
we readily notice that
$$\limgzero k_n(g) = m z_n. \eqno\eq$$
Therefore, for small enough coupling,
the ordinary perturbation theory with normalization adjusted
(such as in \ordresa) is very close to the
improved perturbation theory.
But the question is what happens for $n \rightarrow \infty$
at fixed $g$.
In order to answer this question,
we shall analyse the large order behaviour of $\fn$ in the next
 section.

\section{General formalism for large-order analysis}

We can express $\fn$ as a contour integral in the complex
$g$-plane as,
$$\fn = {1\over 2 \pi i}\ointop {d g\over g^{n+1}} e^{-S[\phi,g]}
      = {1\over 2 \pi i}\ointop {d g\over g}    e^{- \tilde S[\phi,g]},
  \eqn\gcontour$$
where
$$\tilde S[\phi,g] \equiv S[\phi, g] + n \log g .\eqno\eq$$
For large $n$, the saddle point approximation in the
complex $g$-plane is justified.
The saddle points are defined as the solutions of
$${\partial \over \partial g} \tilde S[\phi, g] = 0. \eqn\saddleeq$$
Since the action is usually (at most) quadratic function of $g$,
the above saddle-point equation is a quadratic equation for  $g$, whose
coefficients are functionals of $\phi$.
We shall concentrate on this case hereafter.

Let us now look at the choice of the integration contours.
We denote the action as a function of $g$ as in the following,
$$S[\phi, g] = c_0 - c_1 g + c_2 g^2 .\eqn\definec$$
[The coefficients $c_{0\sim 2}$ are functionals of $\phi$.]
The solutions of the saddle-point equation \saddleeq\ are
$$g_\pm = {c_1 \over 4 c_2}(1\pm \sqrt{D}) , \quad
D \equiv 1-{8n c_2 \over c_1^2} .\eqn\gdefdef$$
The coefficient for the Gaussian integration is
$$ \left. {\partial^2 \over \partial g^2} \tilde S[\phi, g]
    \right|_{g=g_{\pm}}
  = {c_1^2 \over 2n} (-D \pm  \sqrt{D})
      (\equiv \tilde S_{\pm}^{\prime\prime})
  .\eqn\secondderi$$
We have to choose the saddle point(s) and the direction of the
contour to go through it so that the Gaussian integration is
 convergent.
The result is quite different depending on the signature of $D$.
Namely, the whole functional space of $\phi$ is divided into two
parts by the signature of the functional $D$.
We obtain the different expression for $\fn$ in these regions, due
to the difference in the distribution of saddle-points.

For $D>0$, we have a pair of saddle points on the real axis.
Since the functional $c_2$ is positive definite (as we shall see later
in the examples),
$\tilde S^{\prime\prime}$ is negative at $g_-$.
Thus we choose the integration contour to go through the saddle point
 $g_-$
vertically to the real axis \FIG\real{The choice of the contour for
$D>0$ (for $c_1 < 0$).} (see Fig.\real).
As a result, at the leading order of the Gaussian approximation, we
 obtain,
$$\fn = {1 \over \sqrt{2\pi \tilde S_-^{\prime\prime} g_-^2}}
{e^{-S[\phi,g_-]} \over g_-^n} .\eqno\eq$$
In order to find the major contribution (peak) of $\fn$, we need to
 solve for
the following equation,
$${\delta\fn\over\delta\phi(\tau)}=
 \left. {\partial\fn\over\partial g_-}\right|_\phi
  {\delta g_-\over\delta\phi(\tau)}       +
 \left. {\delta\fn\over\delta\phi(\tau)}\right|_{g_-} = 0
 . \eqn\maxcond$$
At the leading order of $\hbar$, however,
the $g_-$ derivative in the first term is zero, simply because of
the saddle-point condition.
Thus we only need to take the second term.
This results in the ordinary equation of the motion,
$$ {\delta S[\phi, g_-]\over\delta\phi(\tau)} = 0. \eqn\ordeq$$
After obtaining a solution $\phi(\tau)$
for a (real) $g_-$, it has to be substituted in the
definition of $g_-$ \gdefdef, which then becomes a self-consistent
 equation
for $g_-$.  Only when one finds a real solution for this
 self-consistent
equation, one finds a peak of $\fn$ in the $D>0$ region of the
functional space of $\phi$.

\def\gabs{|g|}

For $D<0$, the saddle points $g_\pm$ in \gdefdef\ are complex.
We use the following definition,
$$ g_\pm = \gabs e^{\pm i \theta}, \quad
\gabs \equiv \sqrt{n \over 2 c_2}, \quad
\cos\theta \equiv {c_1 \over \sqrt{8 n c_2}}. \eqn\gcomdef$$
(We choose $\theta \in [0, \pi]$, so that $g_+$ is always
in the upper half-plane.)
It is convenient if we rewrite $c_1$ and $c_2$ in terms of
$\gabs$ and $\theta$ using the following relation,
$$c_1 = {2n\over \gabs}\cos\theta, \quad c_2 = {n\over
 2\gabs^2}.\eqno\eq$$
This way we obtain the following expressions at the saddle points
$g_\pm$,
$$S_\pm = c_0 - n - {n\over 2} e^{\pm 2 i \theta}, \quad
\tilde S_\pm^{\prime\prime} = {2n \over \gabs^2} \sin\theta
(\sin\theta \pm i \cos\theta).\eqn\sadexp$$
Since
${\rm Re}(\tilde S_\pm^{\prime\prime}) > 0$
at both saddle points,
\FIG\contour{The integration contour for $g$-integration for $D<0$,
 i.e.,
in case there are two saddle-points which are complex-conjugate of each
 other.}
we choose the integration contour as is illustrated in Fig.\contour.
This results in the following approximation for $\fn$.
$$\fn = {2\over \sqrt{2 \pi}} \ {\rm Im}
 \left[ {1\over \sqrt{S_-^{\prime\prime} g_-^2}}
{e^{-S[\phi, g_-]} \over g_-^n} \right]. \eqno\eq$$
In this case, the equation obtained from \maxcond\ is,
$${\rm Im}\left[ {\delta S[\phi, g_-]\over\delta\phi(\tau)}
 e^{i\sigma}\right]
= 0, \eqn\imeq$$
where $\sigma$ is the phase
$$\sigma \equiv \arg
     \left[ {1\over \sqrt{S_-^{\prime\prime} g_-^2}}
     {e^{-S[\phi, g_-]} \over g_-^n} \right].\eqn\effeq$$
Using \gcomdef\ and \sadexp, we find that
$$\sigma = n\left( \theta - {1\over 2}\sin(2\theta)\right)
  +{\pi\over 4} + {\theta\over 2}. \eqno\eq$$
It is important to note here that in general
the equation \imeq\ is  \undertext{different} from
the equation of motion of the original theory.
This will bring an important consequence later in the analysis.
It also should be noted that although the values of $g$ at the
saddle points are complex, the resulting ``effective" equation of motion

\effeq\ is real.

\chapter{The simple example}

\def\fnr{F_n(\phi)}

In this chapter, we shall revisit the simple example \simple\ in
the previous chapter, using the tools developed so far.

The coefficients $c_i$ defined in \definec\ are;
$$c_0 = {1\over 2} \phi^2, \quad
  c_1 = \phi^3, \quad
  c_2 = {1\over 2} \phi^4. \eqno\eq$$
Note that due to
the symmetry $S(-\phi, -g) = S(\phi, g)$, $F_n(-\phi) = (-1)^n
 F_n(\phi)$.

First, let us look at the region $|\phi | > 2 \sqrt{n}$, where $D>0$.
The solutions of the equation of motion \ordeq\ are
$\phi = 0, 1/(2g_-)$ and $1/g_-$.
The first solution $\phi=0$ corresponds to $D=-\infty$ and therefore
is inconsistent.
The second solution reduces the self-consistent equation \gdefdef\
to the following,
$$g_- = g_- \left( 1 - \sqrt{1-16ng_-^2}\right),\eqno\eq$$
which has the solution,
$g_- = \pm 1/(4\sqrt{n})$.
This corresponds to $\phi = \pm 2\sqrt{n}$,
the boundary of the $D > 0$ region.
At this point,
two saddle points merges,
the second derivative $\tilde S^{\prime\prime}$ vanishes,
and the Gaussian approximation is invalid.
Therefore, this solution is inconsistent.
The third solution $\phi=1/g_-$ leads to
$$g_- = {g_- \over 2} \left( 1 - \sqrt{1-4ng_-^2}\right).\eqn\uninte$$
The only real solution is $g_- = 0$, which corresponds to $\phi = \pm
 \infty$.
This simply corresponds to the fact that $\fn$ is a monotonically
decreasing function for large enough $| \phi |$.
[The corresponding action is $+\infty$ and $\fnr = 0$.]

In the region $\phi < 2\sqrt{n}$ ($D<0$), Eq.\imeq\ becomes
$$\phi \sin\sigma - 3 \gabs \phi^2 \sin(\sigma - \theta)
+ 2 \gabs^2 \phi^3 \sin(\sigma - 2 \theta) = 0. \eqn\physigma$$
It is convenient to scale out $|g|$
using the new variable $\varphi = \gabs \phi$.
This reduces Eq.\gdefdef\ to
$$|g| = \sqrt{\varphi^4\over n}, \eqn\gconst$$
and
$$\cos\theta = {1\over 2 \varphi}.\eqn\thetaconst$$
This way, \imeq\ becomes equation of motion for $\varphi$
where only parameters are $n$ and $\theta$.
Thus, one could first solve it and then determine $\theta$
self-consistently in \thetaconst.
The value of $|g|$ is then determined from \gconst.

\FIG\twinpeaks{The behavior of $\fn$ for the simple model
for (a) $n=7$ and (b) $n=8$.
Note that these are obtained by numerical calculation of
the exact form of $\fn$, not from the saddle-point approximation.}

In the current example, however, it is easier to obtain the
solutions by combining \physigma\ and \thetaconst\ first.
Then we find that the equation reduces to
$$\sin\theta \cos(\sigma + \theta) = 0.\eqn\sigth$$
Among the solutions, $\theta=0$ corresponds to the boundary
of the region, $D=0$, and is inconsistent.
Thus we find $n+2$ solutions labeled by an integer $m=0, ... n+1$ as in
 the
following,
$$\sigma + \theta = \left(m+{1\over 2}\right) \pi.\eqno\eq$$
For $m \ll n$, the solution is approximated by
$$\theta_m = \left({3\pi\over 2}{2m +1 \over 2n}\right)^{1/3}, \quad
\phi_m = 2\sqrt{\pi} - {1\over n^{1/6}}
      \left({3\pi\over 2}{2m +1 \over 2n}\right)^{2/3}, \eqno\eq$$
At these points, the function $\fnr$ is
$$\fnr = {1 \over \sqrt{n}} \left( {3 \over 4\pi}\right)^{1/6}
\left({2n\over 2m+1}\right)^{1/6} n^{n/2} (4e^{-1/2})^n
 \eqn\fnpeakzero$$
The function $\fnr$ is the largest (and positive) for $m=0$.
This and the corresponding peak for $m=n+2$ are the major
contribution to the naive perturbative coefficients $z_n$.
[The solutions for  $m \sim n$ gives the corresponding peaks
for negative $\phi$.]
The function $\fnr$ oscillates quickly-dumped for $|\phi | < \phi_0$.
The solutions $\phi_m$ for $ n+1 \ge m \ge 1$ are the non-leading
 peaks.
This situation can be easily seen in the
numerical plot of $\fnr$ in Fig.\twinpeaks.

After some analysis (following the spirit of Ref.\brezin), we find that

for even $n$,
$$\lim_{n\rightarrow \infty}{z_n \over n^{(n-1)/2} 4^n e^{-n/2}} =
{1\over \sqrt{2\pi}}.\eqn\znbrezin$$
[We have confirmed this result numerically for $n$ as large as 300.]
This result is consistent with our assertion that the peaks at $m=0,
 n+1$
are the major contributions.
Namely, the height of these peaks reproduces the
first two leading terms of \znbrezin,
$$\log z_n \sim {n \over 2} \log n + \left( 2 \log 2 - {1 \over
 2}\right) n
\ (\equiv \log \tilde z_n)
\eqno\eq$$

Let us now look at the improved perturbation theory in this model.
The discrete symmetry we mentioned before is ${\rm Z}_2$,
$$\phi \rightarrow {1\over g} - \phi.\eqn\ztsym$$
The fundamental region we take is
$$\ss = \{ \ \phi \ | \ \phi < {1\over 2g}\ \},\eqn\ztfun$$
which always contains the classical minima $\phi=0$.
The multiplicity $m$ is 2.
Therefore, the coefficient function for the new perturbation series
\zdefimpp\ is defined by
$$k_n (g) = 2 \nn \int\nolimits_{-\infty}^{1/(2g)} d\phi \,
 \fnr.\eqn\goodkn$$

At low orders, we could look at the following expression
obtained from the definitions, \naivezn\ and \goodkn,
$$\eqalign{k_n(g) &=
 2 z_n - {1\over \sqrt{2\pi}} \int\nolimits_{1/(2g)}^\infty d\phi \fnr
 \crr
&\sim
 2 z_n - {1\over \sqrt{2\pi}} \int\nolimits_{1/(2g)}^\infty d\phi
   \ {\ \phi^{3n} \over n!} e^{-{\phi^2 \over 2}} \crr
&\sim
 2 z_n - {1\over n! \sqrt{2\pi}}
    {1 \over (2g)^{3n-1}} e^{-{1 \over 8g^2}} \cr}
      \eqn\loworder$$
This expression is valid for $n \ll 1/g^2$.
Thus the difference between
the improved perturbation theory
\zdefimpp\ and the naive perturbation theory
(with artificial adjustment of the normalization) \ordresa\
is found to be non-perturbative at low orders, $n \ll 1/g^2$.

Let us see what happens for $n \gg 1/g^2$.
We have seen in the above that
the function $\fn$ has two significant peaks at $\phi \sim \pm
 2\sqrt{n}$.
As $n$ increases,
the peak at $\phi \sim 2 \sqrt{n}$ moves out of the integration
region $\phi < 1/(2g)$ for $n > n_c = 1/(16 g^2)$.
On the other hand, the peak at $\phi = - 2 \sqrt{n}$ always remains in
 $\ss$.
The same is true for lesser peaks; the positive ones moves out of $\ss$
for $n\rightarrow\infty$, while the negative one remain.
Therefore, we find that
for $n\rightarrow \infty$ (with $g$ fixed),
$$k_n(g) \rightarrow (-1)^n \tilde z_n .\eqn\impcoeff$$
Namely, the coefficient function $k_n(g)$ is positive for even $n$
and is negative for odd $n$.
Its absolute value
is a smoothly increasing function of $n$ and at even $n$ it is half as
 much as
the naive coefficient.
This series is thus Borel-summable.

\FIG\borel{The value of the partition function ${\cal Z}$
for $g =0.1$ at order $n$ of the perturbative series.
The dash-dotted line is the exact value, the dotted line
the naive perturbation, and the solid line the improved perturbation.}

We have carried out the numerical calculation of
$k_n(g)$ to confirm above analysis.
The results are illustrated in Fig.\borel.
It is apparent that the improved perturbation
theory gives result that oscillates around the
exact value with increasing amplitude,
which is a typical behaviour for asymptotic series,
while the naive result simply diverges at higher orders.\foot{
It is interesting to note that at the transition region
$n \sim n_c$, the original coefficient
$z_n \sim e^{-1/32g^2}$, since $1/32g^2$ is the value
of the action at the peak $\phi=1/2g$, which is the only
analogue of the instanton configuration in this model.}

\chapter{Quantum Mechanics}

We take the following action $S[\phi,g]$,
$$S[\phi,g] = \int\nolimits_0^\beta d\tau \left({1\over 2}\dot\phi^2 +
   {1\over g^2} V(g\phi)  \right), \quad
   V(\phi) = {1\over 2} \phi^2 (1-\phi)^2. \eqn\action$$
The potential is exactly the same as the action in the previous chapter
and therefore many of the features are
quite similar.
This theory has an instanton solution,
$$\phi(\tau) = {1\over g}{1 \over 1 + e^{-\tau}}, \eqno\eq$$
which has the action $1/(6g^2)$.

\section{Evaluation of $\fn$.}

In this case, $c_1 = S_3$ and $c_2 = S_4 /2$, where
$$S_m = \int\nolimits_0^\beta d\tau \, \phi^m. \eqno\eq$$

\def\dt{\delta\tau}

In the region $D > 0$,
a nontrivial solution to the classical equation of motion \ordeq\
with effective coupling $g_-$ is a pair of an instanton and an
 anti-instanton.
This configuration was used in the analysis by Br\'ezin
 et.~al.\refmark\brezin
and is worth looking at.
Let us place the instanton at $\tau_1$ and the anti-instanton at
 $\tau_2$.
We assume that they are well-separated, $1 \ll |\tau_1 - \tau_2| (\equiv
 \dt)$.
This configuration has action $\sim 1/3g_-^2$ and
$$c_1 = {1 \over g_-^3}\left( \dt -3 + ... \right),
\quad c_2 = {1 \over 2 g_-^4}\left( \dt - {11 \over 3} + ... \right).
\eqno\eq$$
(The omitted part is the interaction term between instanton and
anti-instanton,
which decreases with $\dt$ exponentially\Ref\allak{
H.~Aoyama and H.~Kikuchi \journal \PL &247B (90) 75
\journal \PR &43 (91) 1999
\journal {\sl Int.~J.~of Mod.~Phys.} &A7 (92) 2741 (in print).}
and is irrelevant here.)
Using the leading terms,  we find that \gdefdef\ reduces to
$$g_- = {g_- \over 2}\left( 1 -
\sqrt{1 - {4n g_-^2 \over \dt}}\right).\eqn\gmeq$$
The only consistent solution is $g_- = 0$, which
corresponds to $\fn = 0$.
(This conclusion does not change when one includes the non-leading
terms, as long as the instanton and anti-instanton are well-separated
 from
each other.)
As in the case  \uninte\ in the previous chapter, this corresponds to
 the
uninteresting tail of $\fn$.

One may wonder where in our analysis is the instanton
pair that appeared in Br\'ezin et.~al.'s analysis.
It turns out that it corresponds to the ``wrong" saddle point $g_+$.
In fact, if one replaces $g_-$ by $g_+$ in the above analysis, one
obtains
$$g_+ = {g_+ \over 2} {\dt - 3\over \dt -{11\over 3}}
\left( 1 +  \sqrt{1 - 4n g_+^2 { \dt -{11\over 3}\over (\dt - 3)^2}}
 \right),
\eqn\gpeq$$
instead of \gmeq.
For $\dt \gg 1$, this equation has the solution $g_+ = \sqrt{2/3n}$,
which is the coupling found in Ref.\brezin.

Even through $g_+$ is the ``wrong" saddle point for our purpose,
their choice may be perfectly suitable for obtaining the
perturbative coefficient $z_n$.
The coefficient $z_n$ is written as the ``double integral"
$$z_n = {\nn \over 2 \pi i} \intphi \ointop {d g\over g^{n+1}}
 e^{-S[\phi,g]}
    \eqno\eq$$
They choose the instanton-anti-instanton pair as above
and reduce the $\phi$-integral to $\dt$-integration
by integrating over the nontrivial fluctuations.
They integrate over $\dt$ \undertext{along the imaginary axis},
and then finally
carry out the $g$ integration through $g_+$ \undertext{along the real
 axis}.
This procedure was effective in calculating the
$z_n$, but is useless for obtaining $\fn$ or $k_n(g)$.

\def\veff{V_{\rm eff}(\varphi)}
For $D<0$,
the ``effective" equation of motion \imeq\ can be written as
follows by using the scaled variable $\varphi(\tau) = |g| \phi(\tau)$,
$$-{\partial^2\varphi\over \partial\tau^2}
   + {\partial\veff\over\partial\varphi} = 0,\eqn\effeq$$
where the effective potential $\veff$ is defined by,
$$\veff = {1\over 2} \varphi^2 - \kappa_3 \varphi^3 +
   {1\over 2} \kappa_4 \varphi^4, \eqn\veffdef$$
and
$$\kappa_3 \equiv {\sin (\sigma - \theta) \over \sin \theta}, \quad
\kappa_4 \equiv {\sin (\sigma - 2 \theta) \over \sin \theta}.
\eqn\kk$$
The equations \gcomdef\ reduce to
$$|g| = \sqrt{R_4 \over n}, \eqn\gabscon$$
and
$$\cos \theta = {R_3 \over 2 R_4}, \eqn\thcon$$
where we used the following definitions,
$$R_m \equiv \int d\tau \varphi^m. \eqno\eq$$
The equations \gabscon\ and \thcon\ corresponds to \gconst\ and
 \thetaconst\
in the previous example.
The solutions can be now identified by first solving the equation
of motion \effeq, and then by solving the self-consistent equation
\thcon\ for $\theta$.

Let us look at the region $\theta \ll 1$,
since we found the major solution in that region for the simple example

in the previous chapter.
We, however, assume that
$$\sigma \sim {\pi\over 4}+{2\over 3} n \theta^3. \eqno\eq$$
is not necessarily small.
Under this condition,
$$\kappa_3 \sim 1-\theta\cot\sigma, \quad
\kappa_4 \sim 1-2\theta\cot\sigma, \eqn\kk$$
Thus
$\kappa_{3,4}$ diverges at $\theta = \theta_c = (9\pi/8n)^{1/3}$.
We shall look for the solution for $0\le\theta \le \theta_c$.
{}From \kk, we find that the effective potential \veffdef\ is
$$V_{\rm eff} ( \varphi ) = {1 \over 2} \varphi^2 (1-\varphi)
(1 - (1 - 2\theta\cot\sigma )\varphi). \eqno\eq$$
Therefore, for large $\beta$, the solution of equation of motion is
a bounce solution which starts from $\varphi \simeq 0$ and bounces back

at $\varphi \simeq \min(1/\kappa_4,1)$.
Using the \lq\lq energy" conservation law, we can write down
the functions $R_{3,4}$ as integrals
$$R_m = 2 \int\nolimits_0^{\min(1/\kappa_4,1)}
{\varphi^m \ d\varphi
\over \sqrt{2V_{\rm eff} ( \varphi )}}. \eqno\eq$$
Since $\theta \ll 1$,
the self-consistency condition to be solved is now
$$R_3(\kappa_4) = 2 R_4(\kappa_4). \eqn\rtf$$
We have solved this equation numerically and
have found a solution
$\kappa_4 \approx 1.740406$, where
$R_4 \approx 0.143582$.
The value of $\theta$ is then obtained as
$$\theta = \theta_c - {8 \over 9 \pi}
{1 \over \kappa_4 -1}\theta_c^2 + O(\theta_c^3) .\eqn\thsol$$
Thus we can justify the original assumption that
$0\le \theta \le \theta_c$.

For this solution, we obtain the following by using the virial theorem,

$$c_0 = {1\over |g|^2} \left({3\over 2}\kappa_3 R_3 - \kappa_4 R_4
 \right)
 \simeq n(2 - \theta \cot \sigma). \eqno\eq$$
The action is then
$$S \simeq {n\over 2} (\kappa_4 + i \sin 2\theta). \eqno\eq$$
Using this, we find that this solution yields
$$\fn = {1\over\sqrt{\pi n \sin\theta}}
    \left({n\over R_4} \right)^{n/2}
     e^{- {\kappa_4\over 2} n }
     \sin\sigma.
\eqn\fnpeak$$
We have done numerical analysis of $\fn$ in restricted Fourier spaces.
The result confirms that our bounce solution obtained here
is in fact the position of the peak of $\fn$.
The detail is given in the appendix.

We denote this solution by $\phi_+$ and its partner at $\pi - \theta$
 by
$\phi_- (= -\phi_+)$. We find that
$$\eqalign{F_n[\phi_+] &\approx C n^{-2/3} n^{n/2} A^{n/2},\cr
  F_n[\phi_-] &\approx (-1)^n C n^{-2/3} n^{n/2} A^{n/2},}
        \eqn\fnresult$$
where  $C$ is a positive number of $O(1)$ and
$A = e^{-\kappa_4} / R_4 \approx 1.22194$.

Since the leading behavior of the $z_n$ is known,\refmark{\brezin}
one could compare it with the value of $\fn$ at the maxima
\fnresult.
Since $\fn$ is an odd functional of $\phi$ for odd $n$,
$z_n$ is obviously zero.
For even $n$, the known result is
$z_{n} \approx n^{n/2}$.
Thus the most leading term agrees with \fnresult.

\section{Small $g$ approximation}

We shall now discuss the improved perturbation theory.
The action \action\ is symmetric under the global ${\cal Z}_2$,
$\phi(\tau) \leftrightarrow 1/g - \phi(\tau)$.
The fundamental region $\ss$ is then specified by the condition
$$\bar\phi \equiv
  {1 \over \beta}\int\nolimits_0^\beta d\tau \phi(\tau)
     < {1 \over 2g}. \eqno\eq$$

The question is then how $k_n(g)$ would behave.
At the lower order of the perturbation,
$k_n(g)$ differs from $2 z_n$ by only a small amount,
which is non-perturbative in $g$.
This can be shown as using the following relation analogous to
 \loworder,
$$k_n(g) =  2 z_n - 2 \nn \int\nolimits_{\bar\phi > 1/2g}
            {\cal D}\phi \fn \eqno\eq$$
The dominant contribution to the second term
comes from the boundary of the integration
$\bar\phi = 1/(2g)$.
This term can be estimated by
minimizing the free action under the constraint
$\bar \phi = 1/(2g)$.
Using the Lagrange-multiplier method we find that the solution is
the constant one, $\phi = 1/(2g)$, which has the free action $\beta/(8
 g^2)$.
Therefore, the dominant contribution from the second term
is of $O( (\beta/g^2)^n e^{-\beta/(8 g^2)})$.

As $n$ increases, the second term becomes significant.
The solution we have found in the previous section yields
$$\bar \phi_\pm \propto  \pm  {\sqrt{n}\over \beta}, \eqno\eq$$
where the proportionality constant is of $O(1)$.
Therefore, as $n$ exceeds $(\beta / 2 g)^2$, $\phi_+$
moves outside of the integration region
and only $\phi_-$ remains as a major contribution to $k_n(g)$.
For even $n$, this has an effect of reducing the
value of $k_n(g)$ to  $z_n$.
For odd $n$, it yields negative result, which is
the analytic continuation of $z_n$.
In other words, for $n\rightarrow \infty$, the most leading terms are
$$k_n \rightarrow \cases{
   n^{n/2} &for $n=$even,\cr
  -n^{n/2} &for $n=$odd,\cr}\eqno\eq$$
This situation is completely the same as in the
simple example studied in the previous chapter.
Our improved perturbation theory yields a Borel summable series.

\chapter{Discussions}

In this paper, we analyzed the path-integral for theories
that allow tunneling between degenerate
classical minima, which is related by a discrete symmetry.
The non Borel-summability of the perturbation series is
explained as the result of the zero-coupling singularity, where
the degenerate minima move out to infinity and do not contribute to the
path-integral.
For calculations at small but non-zero coupling, we proposed
to do the path-integral only in a fundamental region of the theory.
By calculating the large order behaviour the $n$-th order functional
$\fn$, we have identified the configuration that dominates the
functional integral for the double-well quantum mechanical model.
We have shown that half of these dominant configuration
moves out of the fundamental region for $n \rightarrow \infty$,
leaving the one that contributes with sign $(-1)^n$.
This shows that unlike the naive perturbation theory,
the series expansion we propose is Borel-summable.

Although we have carried out the actual calculation
only for a double-well model in this paper,
the formalism we developed here is applicable for a
wide class of models, including field theoretical ones.

In considering the calculation of the physical observables,
it is important to consider multi-bounce configurations.
In analogy with instanton configuration, it is natural to expect that
there are multi-bounce configuration, which are
not exactly the solution of the equation, but is crucial in the
$\beta \rightarrow \infty$ limit.
Let us take configuration made of $b$ well-isolated bounces.
We assume each bounce is the solution of \effeq.
Since this configuration has $R_{3,4}^{(b)} = b R_{3,4}^{(1)}$,
(we denote the values for $m$-bounces with superscript ${(b)}$),
the self-consistent equations \rtf\
are the same as before.
Thus the same value of $\theta$, \thsol, results.
Then we find that $|g^{(b)}| = \sqrt{b} |g^{(1)}|$.
This means that the height of the each bounce is
$1/\sqrt{b}$ times before.  Due to this, the value of the
total action is the same as before,
$S^{(b)} = S^{(1)}$.
Therefore, for this configuration,
$F_n[\phi^{(b)}] = b^{-n/2} F_n[\phi^{(1)}]$.
Although these configuration has lower value of $\fn$,
the large measure of its collective-coordinate space,
$\beta^b / b!$, makes them important as a whole.

\REF\balitsky{I.~I.~Balitzky and A.~V.~Yung \journal \PL &168B (86) 113
\journal \NP \nextline &B274 (86) 475.}
\REF\shuryak{E.V.Shuryak \journal \NP &B302 (88) 559,621.}
\REF\newvalley{H.~Aoyama and H.~Kikuchi \journal\NP &B369 (92) 219.}

Another interesting feature is the relation with the valley
methods.\refmark{\balitsky - \newvalley}
In these methods and in Ref.\newvalley\ in particular,
the series of configurations (called ``valley") that starts from the
 vacuum
and eventually develops to well-separated instanton and
anti-instanton configuration was identified for the
purpose of carrying out the functional integral
around them.
The bounce-like configuration that dominates $\fn$,
which we found in this paper, is somewhat similar to
intermediate configurations in this valley.
This suggests a possibility that one may split
the functional space into two parts;
the one that contains the outer region of the valley and thus
should be treated by the valley methods and the one
that contains the lower portion of the valley.
The ``improved" perturbation theory for the latter
would be Borel-summable, since the positive bounce
moves out of that region into the non-perturbative region
for $n\rightarrow \infty$.
This may be an appropriate way to treat the $\beta \rightarrow \infty$
 case.
The detailed investigation of these is in progress.

\vskip 1cm
\centerline{\twelvecp Acknowledgements}
One of the authors (H.A.) acknowledges the kind hospitality
of the members of the SLAC theory group, where part of this work was
 done.

\endpage

\appendix

\def\uu{(U^0_n, U^1_n)}
\def\nasp{{n, asymp}}

In this appendix, we present the results of the numerical calculation
for the double-well quantum mechanical model studied in the chapter 4.

In order to evaluate $\fn$ numerically, we first restrict
the functional space of $\phi(\tau)$ to a subspace of finite dimension.
We have chosen to do so by taking a finite number of the
coefficients of the cosine Fourier-expansion of $\phi(\tau)$,
$$\phi(\tau)=\sum_{l=0}^\infty a^l \phi^{(l)}(\tau), \hskip10mm
  \phi^{(l)}(\tau) = \cases{
  \displaystyle{\sqrt{1 \over \beta}}  & for $l=0$ \crr
  \displaystyle{\sqrt{2 \over \beta} \cos \biggl({2 \pi
 l\tau\over\beta}\biggr)}                         &for $l=1,2,3...$ \cr}
                \eqno\eq$$
All the calculation is carried out for $\beta = 20$,
which is large enough to contain
a well-separated instanton-anti-instanton pair.\refmark{\newvalley}

\def\a01{($a^0$, $a^1$)}

First we limit ourselves to the \a01 subspace.
This subspace is small enough to allow us to calculate
exact analytical expression of $F_n(a_0,a_1)$
for $n$ as large as 45 (by using the Mathematica).
\FIG\fourpeaks{Plot of $F_n(a_0,a_1)$ for (a) $n=9$ and (b) $n=10$.}
In Fig.\fourpeaks\ we give the plot of $F_n(a_0,a_1)$
for odd $n$ and even $n$.
It is apparent that  $F_n(a_0,a_1)$ has well defined peaks.
We have calculated the position and the height of these peaks
 numerically.

On the other hand, in our $n \rightarrow \infty$
asymptotic analysis, we can determine \a01 and $\kappa_{3,4}$
as follows:
First we substitute
$$\varphi(\tau)= \sum_{l=0}^\infty \alpha^l \phi^{(l)}(\tau)
\quad (\alpha^l \equiv \gabs a^l), \eqno\eq$$
into the effective equation of motion $\effeq$.
We obtain two equations from the first two modes of this equation.
The other two equations we need are the self-consistency condition
 $\rtf$
and the relation $2\kappa_3=\kappa_4+1$, which comes from \kk.
We have found only one non-trivial real positive answer, which is
$\alpha^0 = 1.16625$ and $\alpha^1 = 0.978323$.
This corresponds to $R_4 = 0.551747$ and $|g| = 0.742797/\sqrt{n}$.
This leads to
$$ (a_\nasp^0,a_\nasp^1)=(1.57008 \sqrt{n},1.31708 \sqrt{n})
 \eqn\azoasym$$

Let us compare the exact and asymptotic values.
In general we expect to see that the exact result and the asymptotic
 one
agree only in the leading terms, which is O($\sqrt{n}$).
In order to see it clearly, we consider a pair of quantities
$$ ( U^0_n , U^1_n ) =
     2 \sqrt{ n } ( a^0_n - a^0_{n-1} , a^1_n - a^1_{n-1} )
                                                         \eqn\uudef$$
and similarly for the asymptotic answer.
This way, if the next-leading term is a constant,
it is subtracted out of the above expression
and the exact $( U^0_n , U^1_n )$ should converge to the asymptotic
 value
$$ ( U^0_\nasp , U^1_\nasp ) = (1.57008,1.31708)
\eqn\upoint$$
obtained from \azoasym.

\FIG\ffst{The point $\uu$ for every fifth $n$ from 5 to 40 is plotted.
The point marked by a cross is the asymptotic answer \upoint.
Good convergence for $n \rightarrow \infty$ is apparent.}

\FIG\fsnd{Plot of $V_n$ and $V_\nasp$ defined in \vdef.
These seemingly converge to a same value, which is a supporting evidence
 for
the agreement of the exact result and the asymptotic one,
$\gamma=0$ and $A=A'$.}

\FIG\ftrd{The shape of $\varphi(\tau)$ at the peak of $F_{10}[\phi]$.
The solid lines are for the peaks we found in the restricted subspaces,

$m = 1 \sim 4$, in the ascending order of the height of the peaks.
The dotted line is for the bounce solution we have found.}

\FIG\ftth{The shape of $\varphi(\tau)$ at the peak of $F_n[\phi]$
in the five dimensional subspace, $m=4$.
The solid lines are for $n=1,4,7$ and 10
in the ascending order of the height of the peaks.
The dotted line is for the asymptotic answer.
We observe that the solid lines approach to the dotted line as $n$
 increases.}

In Fig.\ffst, we plot $(U^0_n,U^1_n)$ for every fifth $n$
as large as 45 and the point \upoint.
We find that the convergence is in fact excellent.
It is apparent that the difference goes to zero
in proportion to $1/n$.
Thus the difference between $a_n$ and $a_\nasp$
is of $O(n^0)$.

We next discuss the height of the peak of $\fn$.
In general we assume that $F_n(a^0_n,a^1_n)$ have the form,
$$ F_n( a^0_n , a^1_n ) = C n^\delta n^{{n \over 2} + n \gamma}
                         A'^{{n \over 2}},
 \eqn\fnex$$by an analogy with that of the asymptotic answer,
 $\fnresult$.
Here $C$, $A'$, $\delta$ and $\gamma$ are constants.
To see the convergency explicitly,
we define $\widetilde F_n$ as $F_n$ divided by $n^{n/2}$ and consider
$$ V_n = {\widetilde F_{n+5} \over \widetilde F_n}.         \eqn\vdef$$
If we substitute $\fnex$ into the above, we obtain the following,
$$ V_n = (n+5)^{5\gamma} \left( 1+{5 \over n} \right)^{n \gamma +
 \delta}A'^{{5 \over 2}}.
 \eqn\vex$$
As $n \rightarrow \infty$, the right hand side diverges if $\gamma>0$
 and goes to zero if $\gamma<0$.
In Fig.\fsnd, we find that neither happens; $\gamma=0$.
Then the right hand side of $\vex$ becomes
$$ V_n \simeq A'^{{5 \over 2}} + O \left( {1 \over n} \right)
        \eqno\eq$$
for large $n$.
On the other hand, for large $n$, $V_\nasp$ is expressed as
$$V_\nasp \simeq A^{{5 \over 2}} + O \left( {1 \over n} \right)
           \eqno\eq$$
The behaviour of $V_n$ and $V_\nasp$ in Fig.\fsnd\
is consistent with their convergence, which means $A=A'$.

We have also investigated the peak structure in $F_{10}[\phi]$
in a series of subspaces ($a_0, ... , a_m$) for $m = 1 \sim 4$.
Namely, we obtained analytical exact expression
of $F_{10}(a_0, ... , a_m)$ and calculated the position of the
peak numerically.
The shape of the function $\varphi(\tau)$  at the peak is
drawn in Fig.\ftrd\ for each case.
Also drawn is the shape of the bounce solution $\varphi_+$
obtained in the main body of this paper
(which, of course, contains all the Fourier modes).
It is apparent that as the subspace is expanded,
the shape of $\phi(\tau)$ approaches
that of the asymptotic answer.

Finally, we consider the five dimensional subspace ($m=4$)
and vary $n$ from one to ten
(which is the practical upper limit of our computational power).
In Fig.$\ftth$ we give the shape of the function $\varphi(\tau)$
at the peak of $F_n[\phi]$ for $n=1,4,7$ and $10$.
We can see clearly that the approximation by the asymptotic answer
 becomes
better as $n$ gets larger.

In conclusion,
the numerical calculation we performed strongly supports
the dominance of the bounce solution in $\fn$ and
the validity of analytical methods we developed in this paper.

\endpage

\refout
\endpage
\figout
\bye